# Integrating Experiment with Theory to Determine the Structure of Electrode-Electrolyte Interfaces

Lalith Krishna Samanth Bonagiri[1,2], Amir Farokh Payam[3], Narayana R. Aluru[4], Yingjie Zhang[1,5,6]*

1. Materials Research Laboratory, University of Illinois, Urbana, Illinois 61801, United States

2. Department of Mechanical Science and Engineering, University of Illinois, Urbana, Illinois 61801, United States

3. Nanotechnology and Integrated Bioengineering Centre (NIBEC), School of Engineering, Ulster University, Belfast BT15 1AP, United Kingdom

4. Walker Department of Mechanical Engineering and Oden Institute for Computational Engineering & Sciences, The University of Texas at Austin, Austin, Texas 78712, United States

5. Department of Materials Science and Engineering, University of Illinois, Urbana, Illinois 61801, United States

6. Beckman Institute for Advanced Science and Technology, University of Illinois, Urbana, Illinois 61801, United States

*Correspondence to: yjz@illinois.edu

**Abstract:** Electrode-electrolyte interfaces are crucial for electrochemical energy conversion and storage. At these interfaces, the liquid electrolytes form electrical double layers (EDLs). However, despite more than a century of active research, the fundamental structure of EDLs remains elusive to date. Experimental characterization and theoretical calculations have both provided insights, yet each method by itself only offers incomplete or inexact information of the multifaceted EDL structure. Here we provide a survey of the mainstream approaches for EDL quantification, with a particular focus on the emerging 3D atomic force microscopy (3D-AFM) imaging which provides real-space atomic-scale EDL structures. To overcome the existing limits of EDL characterization methods, we propose a new approach to integrate 3D-AFM with classical molecular dynamics (MD) simulation, to enable realistic, precise, and high-throughput determination and prediction of EDL structures. As examples of real-world application, we will discuss the feasibility of using this joint experiment-theory method to unravel the EDL structure at various carbon-based electrodes for supercapacitors, batteries, and electrocatalysis. Looking forward, we believe 3D-AFM, future versions of scanning probe microscopy, and their integration with theory offer promising platforms to profile liquid structures in many electrochemical systems.

**Table of Contents Graphic**

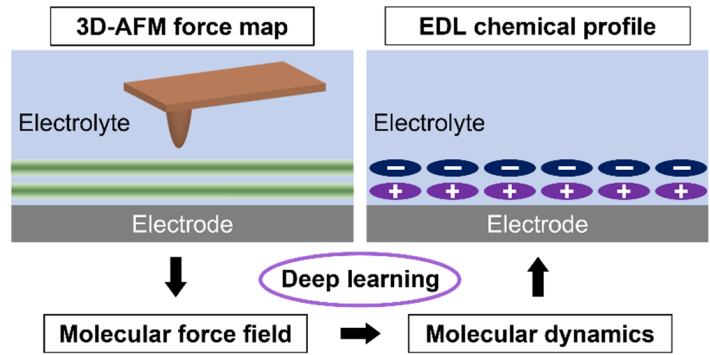

# 1. Introduction

Solid-liquid interfaces are ubiquitous in various natural and engineered systems, and are critical for a large variety of functions, such as biomolecular signal transduction,[1–3] water filtration,[4–7] corrosion control,[8–10] and electrochemical energy conversion and storage.[11–18] At the transition region between the solid surface and the bulk liquid, the interfacial molecules form solvation layers, also called the electric double layers (EDLs). Such interfacial layers play pivotal roles in various electrochemical processes. For example, in supercapacitors, molecular rearrangements in the EDL is the underlying principle for energy storage;[19–21] in electrocatalysis and fuel cells, interfacial solvation strongly modulates the configuration and binding energy of reactant and intermediate species, thus regulating the kinetics of the electrochemical reactions;[22–25] in lithium-ion batteries, $Li^+$ solvation in the EDL has been widely assumed to be a key factor determining the anode passivation, energy density, and battery lifetime.[18,26–28] However, due to the complex interaction with both the rigid solid structure and the mobile liquid species, the molecular force fields in EDL are highly convoluted and challenging for atomistic simulations. Meanwhile, the buried nature and incompatibility with vacuum environments also pose obstacles for experimental characterization of EDLs. As a result, despite decades of research efforts, the fundamental EDL structure and interaction mechanisms at solid-liquid interfaces remain elusive.

Existing experimental studies of solid-liquid interfaces mainly use either spectroscopy methods, such as X-ray and optical spectroscopy,[29–36] or imaging techniques such as electron and scanning probe microscopy.[37–49] Each of these existing methods provide only partial information of the atomistic interfacial structure, such as the overall chemical bonding states, the structure of solid surfaces, or the vertical force profiles of EDL. On the other hand, atomistic simulations, such as density functional theory (DFT) and MD,[50–56] depend sensitively on the choice of approximate functions and parameters such as density functionals, force fields, and many-body interaction terms. As a result, existing simulation methods have not been able to accurately predict the realistic structure of solid-liquid interfaces. Moreover, existing experimental and theoretical approaches have been largely independent, with no synergistic feedback between each other. As a consequence, studies of interfacial structure using different methods tend to show discrepancies. Therefore, an experimental method to thoroughly determine the 3D atomistic interfacial structure, as well as the full integration of experimental and theoretical approaches, are critically needed to bridge the knowledge gap of solid-liquid interfaces.

In this perspective, we first discuss the experimental methods for in situ characterization of electrode-electrolyte interfaces and their pros and cons, followed by a brief review of the widely used MD simulation methods. We will then propose a method for integrating experiment with simulation, and eventually discuss the potential applications of such integrated platform in real-world electrochemical systems.

# 2. Experimental Characterization of EDLs

**2.1 Overview of the In Situ Methods to Characterize Electrode-Electrolyte Interfaces.** Due to the structural complication of liquid electrolytes, existing approaches for in-situ characterization of the electrode-electrolyte interface have achieved limited success. Typical challenges include low resolution, limited sensitivity, and perturbations to the inherent solid-liquid interfacial structure, as described in the following paragraphs.

Spectroscopy methods, such as X-ray, Raman, infrared, and vibrational sum-frequency generation spectroscopies, have been widely used to probe the chemical nature and bonding states of interfacial species.[29–36] Surface sensitivity can be achieved through various mechanisms such as the short electron inelastic mean free path, plasmonic enhancement, total internal reflection, and inversion symmetry breaking. However, the exact thickness of the interfacial region sensed by each method is oftentimes unknown, and the in-plane spatial resolution is either absent or limited to the micron scale.

X-ray scattering, powerful for determining solid structures via Fourier transform, has also been used to extract EDL distribution near solid surfaces.[23,57] However, due to the weak ordering and small thickness of these interfacial liquid regions, these scattering-based methods are restricted to single crystal model electrodes, and the signal is usually dominated by the electrode that has more crystalline ordering than the EDL. Although the vertical distance between the first adsorbed layer and the electrode surface can be indirectly inferred from fitting analysis, the in-plane spatial distribution, and the upper layers of ions/molecules in the EDL cannot be resolved.

Liquid phase electron microscopy is another method that is powerful for imaging solid structures immersed inside liquid environment.[38] The dynamic structural evolutions of nanoparticles have been determined using this technique.[58] Nevertheless, the electron dosage inherently affects the interfacial electrochemical reactions, especially when high dosage is used for atomic resolution imaging.[38] Moreover, this technique currently cannot resolve the molecular distribution in EDL which is a key component of the solid-liquid interface, and the images are typically convoluted by a combination of bulk and surface solid structure.[37]

Electrochemical scanning tunneling microscopy (EC-STM) can directly resolve the surface atomic structure of the solid electrodes, but cannot image the full molecular distribution in EDL, since the small tunneling distance (typically a few angstroms) between the tip and solid surface can lead to the repelling of molecules before they can be detected.[42–44] As a result, EC-STM can only sense the presence of strongly adsorbed molecular species on the electrode surface, which is only a small fraction of the EDL. For example, 1-ethyl-3-methylimidazolium bis(trifluoromethylsulfonyl)imide (EMIM-TFSI) is a widely used ionic liquid for energy conversion/storage applications. However, it was shown that EC-STM cannot observe any feature beyond the graphite lattice at the EMIM-TFSI / graphite interface, and molecules can only be observed when a much larger cation, $C_8MIM$, is used to replace the EMIM.[42] In addition to a lack of molecular-imaging capabilities, EC-STM also tends to have strong perturbation effects, since the metallic tip, within a few angstroms away from the electrode surface, can dramatically modify the local electric field and disturb the interfacial electrochemical processes.[59,60]

AFM is another tool that has been used to characterize solid-liquid interfaces. As an imaging technique based on force detection,[61,62] AFM does not require the transmission or conduction of electrons, and uses electrically insulating tips, thus largely preserving the native electrochemical environment. Previous studies on ionic liquid/electrode interfaces have reported discrete spikes in force-distance curves, revealing the presence of individual molecular layers.[39–41,45–49] In the past decade, 3D-AFM has been implemented by a few groups worldwide, and has been used to image the EDL of aqueous solutions.[63–71] Our lab recently developed a novel electrochemical 3D-AFM (EC-3D-AFM) method that enables direct imaging of the 3D atomic-scale structure of electrode-electrolyte interfaces at controlled electrode potential. As shown in our recent publications[72–74] and Figure 1, we have obtained the atomically resolved structure of various solid surfaces, 3D molecular density maps of the EDLs of a diverse set of electrolytes, as well as the reconfiguration of EDLs at different electrode potentials. As a force mapping method, our EC-3D-AFM is capable of directly resolving the spatial configuration of the immobile solid surface (including deposited clusters/films). For the mobile species in the EDL, quantitative analysis of the force map will enable the determination of the local density distribution, characteristic size of the solvation clusters, inter-molecular/inter-cluster interaction strength, and the pair correlation function of the dominant molecular/cluster species.

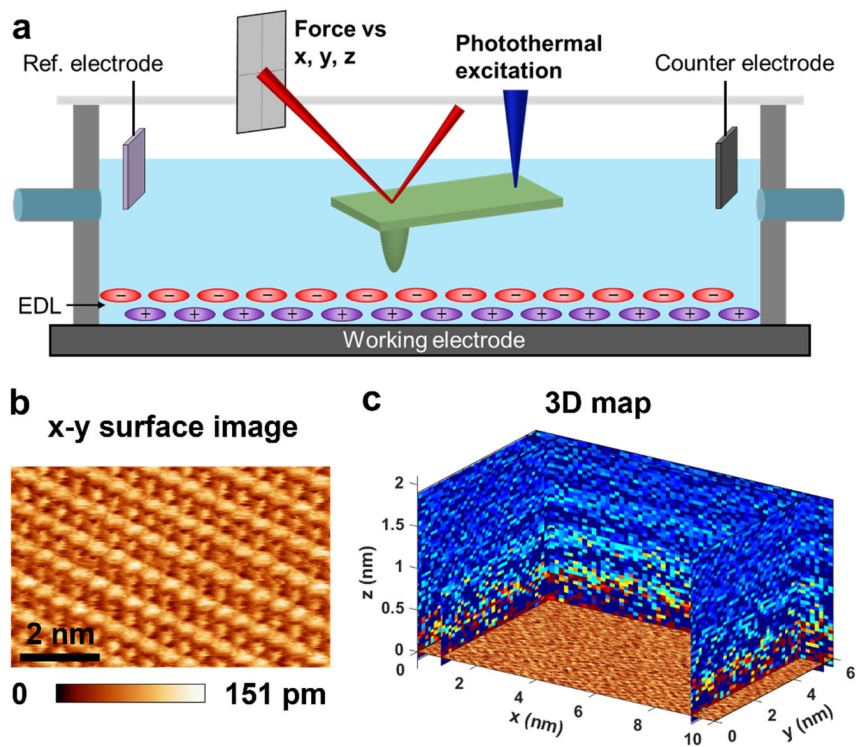

**Figure 1.** In situ EC-3D-AFM characterization. (a) Schematic of our EC-3D-AFM setup. (b) An atomic-resolution x-y image of calcite surface in water. (c) 3D image of the EDL of EMIM-TFSI on a crystalline surface of highly oriented pyrolytic graphite (HOPG).[72]

**2.2 Technical Details of EC-3D-AFM.** Our EC-3D-AFM integrates three methods to enable atomic-scale imaging under controlled electrochemical conditions:

(i) Photothermal excitation. Traditional liquid-phase AFMs use acoustic excitation to drive cantilever oscillation and detect the tip-sample interaction. This induces mechanical oscillations of not only the micro-cantilever, but also other parts of the AFM tip holder, which strongly couples to the hydrodynamic motion of the fluid leading to large mechanical noise.[75,76] To overcome these limitations, we have utilized a photothermal excitation method, which uses an intensity-modulated laser to induce periodic temperature variation of a micro-spot on the cantilever (Figure 1a), and excite smooth mechanical oscillation of only the cantilever itself. Using this technique, we achieve down to ~1 pN force sensitivity, and sub-10 pm spatial resolution. Considering that the inter molecular / atomic distances and tip-molecule interaction forces are typically at least 100s of pm and 10s of pN, this EC-3D-AFM method is capable of resolving the individual atoms and molecules at the solid-liquid interfaces.

(ii) 3D high-speed force mapping. At room temperature, there is always a finite thermal drift of the tip and sample position. In order to obtain accurate 3D molecular images (e.g. in a 10 × 10 × 5 nm$^3$ area), the data acquisition needs to take place within ~1 min, requiring a z rate of ~100 Hz or higher, which is beyond the capability of typical AFMs. We mitigated this problem by implementing a high-speed force mapping technique that uses a sinusoidal wave to drive the tip motion in the z direction with a frequency up to 1 kHz, while simultaneously moving the tip along x and y directions. This sinusoidal wave overcomes the limits of the typically used triangular waves, which generate noise at turning points when operating at high speeds.

(iii) Sealed EC cell with liquid and gas perfusion. We have implemented an EC cell for the 3D-AFM (Figures 1 and 2), which includes a working electrode (the sample), a counter electrode (e.g., Pt, Cu, or Li), and a reference electrode (e.g., Pt (quasi-reference) or Li for organic electrolytes, and Ag/AgCl for aqueous solutions). The EC cell is fully sealed except for ports for liquid and gas perfusion. These ports can be either sealed or used to replace the liquid electrolyte and/or the gas environment inside the EC cell. Moreover, all the parts of the probe holder are fully electrically insulating, thus minimizing any possible electrostatic perturbation effects. This design allows us to perform stable, continuous 3D-AFM imaging under controlled gas atmosphere, varying electrolyte composition, and precisely modulated electrode potential, all of which are critical for in-situ observation of electrochemical processes at solid-liquid interfaces.

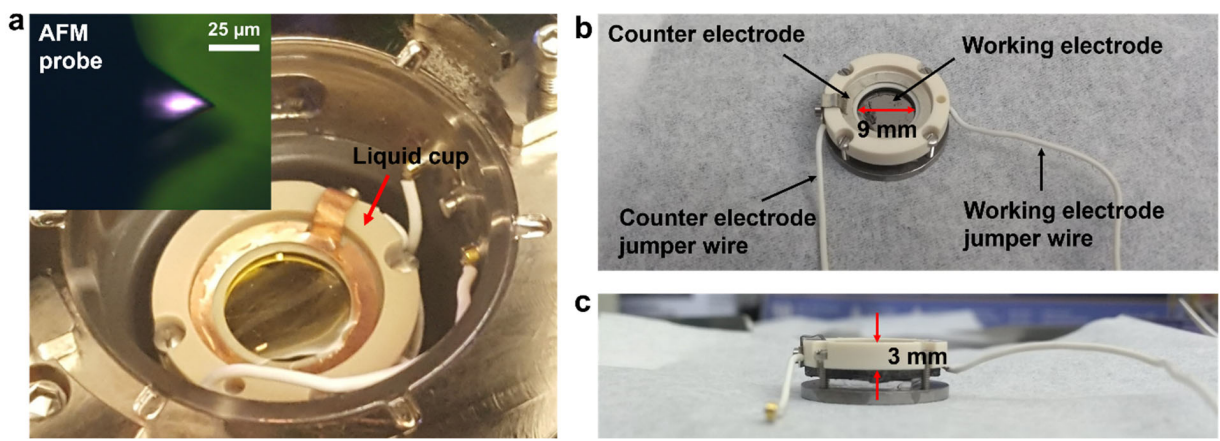

**Figure 2.** Photos of the EC cell for EC-3D-AFM.[72] (a) Photo of our AFM EC cell (open) and an AFM cantilever immersed in liquid (inset). The EC cell will be covered and sealed by a probe holder during imaging. (b, c) Photos of the liquid cup, a key part of the EC cell shown in (a).

Based on our recent series of work on EC-3D-AFM and our prior experience on scanning probe instrumentation and imaging in air,[77–94] we believe it is important to clarify the differences between air and liquid imaging as well as the critical factors responsible for atomic/molecular resolution in liquid. It is our hope that the awareness of these factors in the materials, chemical, and biological science communities will enable the broader and more effective usage of liquid-phase or electrochemical AFM in either the surface or 3D imaging mode. The key atomistic imaging mechanism, non-invasiveness, and protocols to ensure data reproducibility of the EC-3D-AFM method are summarized below.

(i) The effect of tip radius. A common misconception on atomic-resolution AFM is that the tip needs to be super sharp (e.g., ~1 nm tip radius). On the contrary, it has been shown that atomic resolution can be routinely achieved using tips with radii between 5–30 nm, and the image quality of crystalline surfaces does not rely on the value of the tip radius.[95] This is because atomic-resolution imaging is based on the short-range force that occurs between the one or few atoms at the very end of the tip and the atoms/molecules on the sample surface. The overall nanoscale curvature does not contribute to the short-range tip-sample interaction and thus has negligible impact on the ultimate resolution limit.

(ii) Tip perturbation effects. Another typical concern on AFM imaging is the possible perturbation of the tip to the intrinsic materials structure. In our EC-3D-AFM this effect is mostly negligible, due to two factors. (1) Reversible recovery. During 3D-AFM, images are obtained as the tip constantly pushes through the EDL and senses the interaction force. While the local molecules are temporarily repelled away from the area occupied by the tip, they can recover their original position right after the tip retracts from the local spot (within pico- to nano-seconds).[66] Therefore, the captured force profile will be reproducible and genuinely represent the actual EDL structure. Our results reveal that phase/force curves continuously measured at the same spots exhibit the same discrete molecular layered structure (Figure 3), confirming the reversible recovery effects. (2) Tip insulation. Most of our AFM tips have a surface composition of $SiO_2$ or $Si_3N_4$. Both the tips and the probe holder are electrically insulating, which ensures minimum electrostatic influences on the EDL structure. It has been previously shown that silicon nitride tips and Au-coated tips, if electrically floating, produce the same inter-layer spacing when imaging the EDL of ionic liquids, confirming that electrostatic perturbations are negligible.[96]

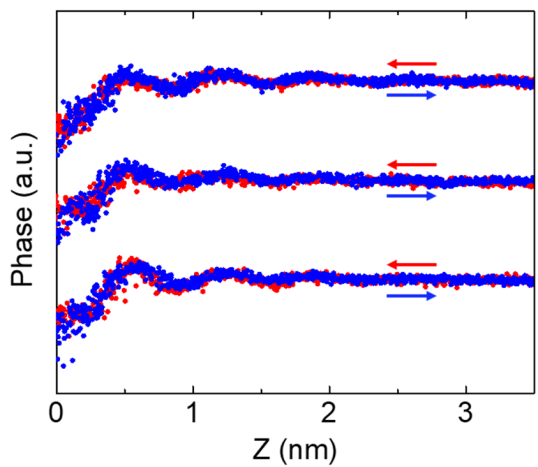

**Figure 3.** AFM phase vs distance curves measured at the same spot in an ionic liquid, EMIM-TFSI, on HOPG.

(iii) Tip cleanliness. A key factor affecting the AFM image resolution and reproducibility is the tip cleanliness.[95,97] This is because the typical organic contaminants on the tip can both induce a large long-range attraction with the sample that overshadows the short-range force, and can constantly deform and result in imaging artefacts. We have established two protocols to ensure tip cleanliness and data reproducibility: (i) tip pre-cleaning using solvents and UV Ozone, to remove organic contaminants right before each EC-3D-AFM experiment; and (ii) tip condition monitoring by constantly performing force-distance curves on fixed spots during the EC-3D-AFM measurements and examining the long-range force profile. We are usually able to continuously image in liquid for at least two days without observing obvious tip degradation effects, which is enough to thoroughly capture the electrochemical processes of one electrode sample.

## 3. MD Simulation of EDLs

MD simulations have been widely used to study many molecular and complex systems, as recognized by the 2013 Nobel Prize in Chemistry. In this perspective, it is not our intention to give a thorough review of MD simulations. Rather, we will focus on the application of MD to investigate EDL structures. Mainly two types of MD simulations have been employed to study electrode-electrolyte interfaces: ab initio and classical MD.

Ab initio MD (AIMD) calculates the energy and force "on the fly" from first principles and then uses such a force to compute atomic motions through classical mechanics. Due to the high computational cost, to date AIMD is still mainly limited to small systems with a size no more than a few nanometers and a time scale within ~100 ps. For EDLs in particular, AIMD has been used to predict liquid electrolytes with small solvent molecules such as water. However, even for these seemingly simple systems, the simulation results still do not fully agree with experimental observations.[29,98] For more complex electrolytes such as ionic liquids, the inaccuracy of AIMD becomes larger.[99]

Classical MD (CMD) works by numerically solving Newton's equation of motion to determine the trajectory of atoms and molecules in a system. Since the motion of atoms and molecules is calculated based on the force field (FF) inputs, the ultimate precision of CMD simulations depends on the choice of FF parameters. Due to its relative simplicity, CMD is much less computationally demanding compared to AIMD and can enable all-atom simulations in systems of at least ~100 nm in size.[100]

In our prior research, we have performed extensive CMD simulations to understand interfacial structure of water and ions near various surfaces including silicon,[101–104] silica,[105] silicon nitride,[106] carbon nanotubes,[107] graphite,[72] etc. The FFs needed in these MD simulations are extracted from detailed quantum calculations and are summarized in Refs. [108,109] for graphite/graphene, Refs. [110,111] for hexagonal boron nitride, and Ref. [112] for $MoS_2$. To mimic the experimentally applied electrode potential, various charge densities are assigned to the electrode surface for MD simulation. After the simulation equilibrates, the Poisson equation is solved using the charge density distribution in the EDL, and then the electrode potential is calculated.[72]

To enable direct comparison between CMD and 3D-AFM results, we have studied a model system, EMIM-TFSI on HOPG. We have further extended our EC-3D-AFM capabilities to perform charge profiling 3D-AFM (CP-3D-AFM), which combines force maps with electrostatic solver to determine the real-space charge distribution.[74] As shown in Figure 4, our CMD and CP-3D-AFM results exhibit qualitative agreements in charge distribution, both revealing that the nearest EDL layer (with ~1 nm from the HOPG surface) dominates the capacitive charging response, and that the cations and anions move in opposite directions as we change the electrode potential/charge density, due to electrostatic modulation. However, quantitatively we observe discrepancy of the z position of the charges. Although CP-3D-AFM may have lower spatial resolution than CMD, such discrepancies in the peak positions still exist even if we apply broadening to the MD density peaks (to mimic lower resolution). Therefore, it is likely that the accuracy of CMD is still limited and the highly popular All-Atom Optimized Potential for Liquid Simulations (OPLS-AA) force field[113,114] that we used can be further improved.

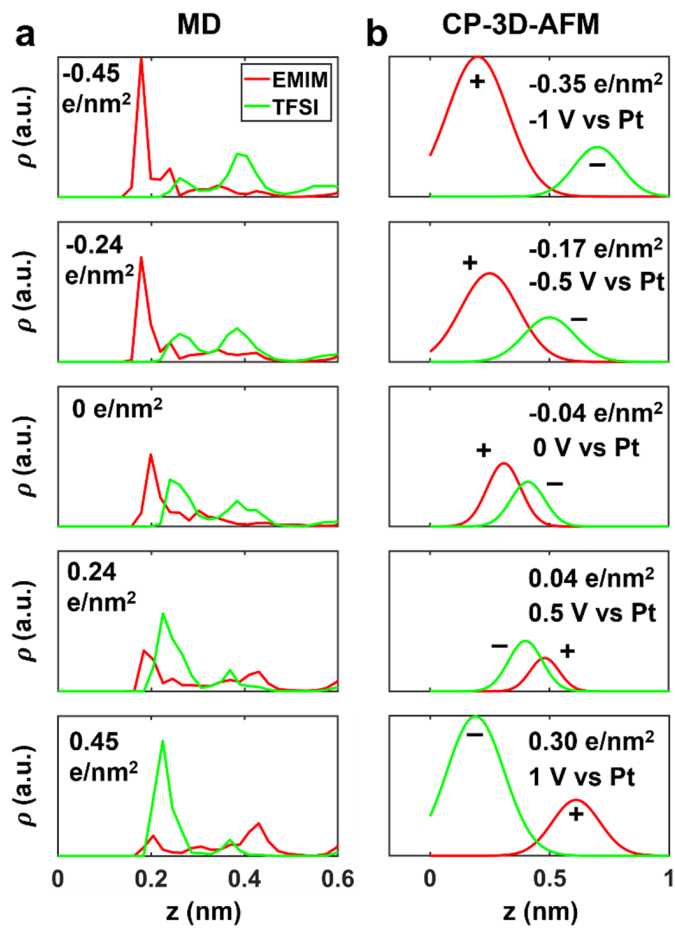

**Figure 4.** Comparison of the charge density vs z profiles of EMIM-TFSI/HOPG from MD simulation and CP-3D-AFM experimental results.[74]

## 4. Prospects for Experiment-Theory Integration to Determine the EDL Structure

Considering that both the experimental characterization and numerical simulation methods have limitations in determining the precise EDL structure, it is desirable to find ways to integrate experiment with theory to improve the accuracy. To this end, 3D-AFM force maps are ideal for integration into simulations due to the high, valuable information density contained in the 3D force vs x, y, z data. However, to integrate 3D-AFM with CMD, we need to first bridge the resolution gap, since the limited experimental resolution does not match with the infinitely small spatial grid of the MD results, as evident in Figure 4. In addition, to date the charge quantification capabilities of CP-3D-AFM can only be applied to highly ionic electrolytes; for dilute solutions (e.g., aqueous solution with <1 M salt concentration), it has not been possible to extract quantitative molecular or charge density profiles from the force maps, posing an "information gap" for integrating with

MD.

We propose one approach to bridge the gap between CMD and 3D-AFM: include an "AFM tip" in the MD simulation of an electrode-electrolyte interface system. Such a virtual tip can be constructed as a nanocluster having the same composition as the real AFM probe used in 3D-AFM imaging.[115,116] This nanocluster can be fully immersed in the liquid electrolyte. At each probe position, an MD simulation can be run to reach equilibrium, when the overall force impinged on the probe can be extracted. Such "virtual AFM imaging" in MD will produce a 3D force map, which, after normalization, can be directly compared to the experimental 3D-AFM force map.

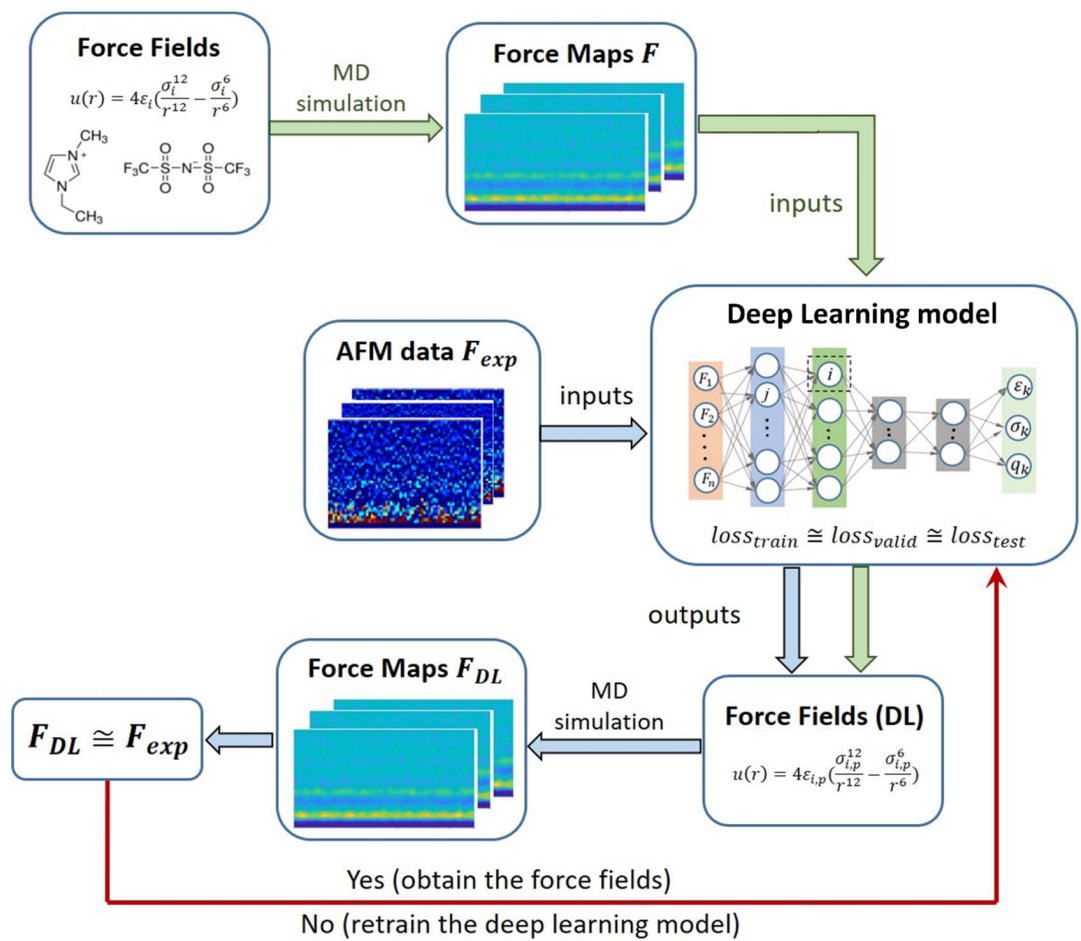

**Figure 5.** Schematic of a feasible deep learning (DL) method for integrating 3D-AFM data with MD simulation.

To integrate the 3D-AFM results into CMD, one can employ a deep learning approach, with an example shown in Figure 5. A deep neural network (DNN) model can be developed to relate the force maps (inputs) to the FFs (outputs). Since DNN requires large datasets, one may generate the datasets using MD simulations. The simulated systems can be built by incorporating the virtual AFM tip into the liquid electrolyte, and then the force maps can be directly extracted from the simulations. Since the force maps depend on the FFs employed, one may generate large datasets of force maps by varying the FFs. The dataset can then be separated into training, validation, and testing dataset. The training dataset can be used to train the DNN model to relate the force maps to FF parameters. The DNN architecture may need to be redesigned (as necessary) until the model can draw accurate correlations between force maps and FF parameters in the validation and testing datasets. Once the training process is accomplished, the force maps obtained from 3D-AFM experiments can be used as inputs to the deep learning model to determine the related FFs. These FFs may subsequently be used in MD to compute the force maps, and the resulting MD force maps can be compared with the experimental force maps. When the deviation is negligible, the estimated FFs are accurate, and MD simulations (with no virtual AFM tip present) can then be performed again to determine the precise structural and dynamical properties of the EDL. If the deviation is noticeable, the deep learning model will need to be further trained by incorporating more training datasets until the error between the experimental and MD force maps is negligible.

## 5. Application Examples

The integration of atomic-resolution EC-3D-AFM and MD simulations will open up many new directions for the precise determination and prediction of electrode-electrolyte interfaces. As an extension of our existing work on HOPG/electrolyte interfaces, we will discuss about prospects of EDL structure at more heterogeneous carbon electrodes, as well as their impact on real-world applications in batteries, supercapacitors, and electrocatalysis.

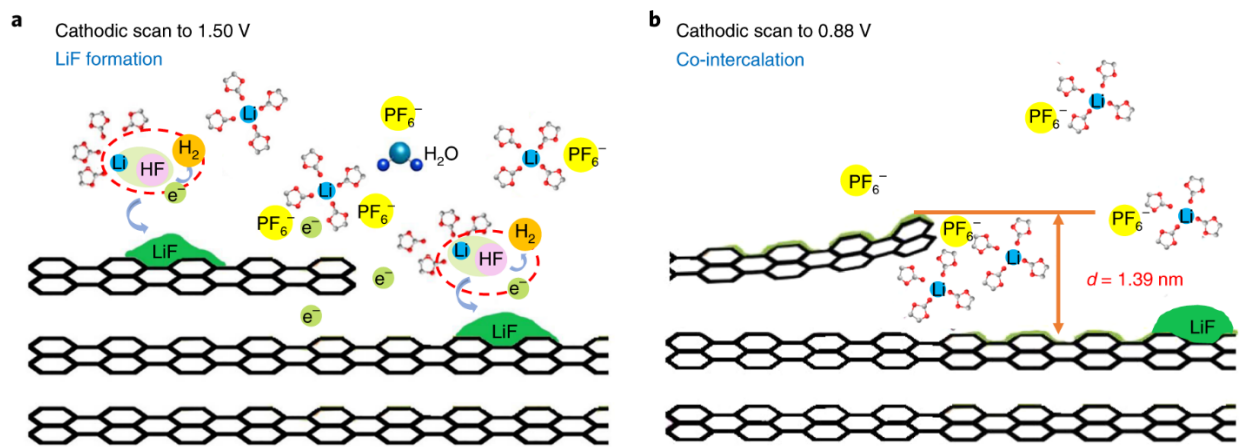

**Figure 6.** Schematic illustration of the initial lithiation process at graphite anode in 1 M LiPF$_6$ EC/DMC electrolyte.[27]

**5.1 EDL at HOPG step edges.** Step edges occur ubiquitously in all types of carbon electrodes. In supercapacitors, for example, the widely used activated carbon and other types of nanoporous carbon electrodes inevitably have significant amounts of exposed edge sites at the atomic scale.[117] In lithium-ion batteries, the graphite anodes usually have edge planes exposed to facilitate ion intercalation.[27,118] Figure 6 illustrates a possible interfacial structure during the initial charging/lithiation process in a commonly used lithium-ion battery electrolyte, 1 M LiPF$_6$ in ethylene carbonate / dimethyl carbonate (EC/DMC).[27] Although it has been widely hypothesized that Li$^+$ ions remain solvated during the intercalation process,[119,120] MD simulations have revealed that the Li$^+$ solvation shell at graphite/electrolyte interface can be dramatically different from that of the bulk electrolyte,[121] posing a question on the exact $Li^+$ solvation state at the edge intercalation site. The atomic-resolution EC-3D-AFM will shed light on the liquid density distribution at such step edge sites, and once integrated with MD, will enable the precise determination of the local $Li^+$ solvation structure. Such results would be transformative in battery research, as it will likely enable the predictive design of electrolytes to promote favorable $Li^+$ intercalation processes for faster charging and/or higher energy density.

**5.2 EDL at defect and dopant sites of carbon electrodes.** In carbon electrodes, miscellaneous defect sites are inevitable in real-world application, especially in supercapacitor systems where vacancies/pores are oftentimes produced during the processing of various nanostructures that are used to achieve large surface areas.[122–124] An atomistic picture of the EDL structure near such sites will be critical for understanding the overall capacitive energy storage mechanism. As a mechanistic study, in order to controllably introduce vacancies/pores without generating other types of defects, one may perform low-power argon plasma treatment to HOPG, which knocks off atoms via physical ion bombardment effects. These vacancies can have different sizes and topography, ranging from single-atomic vacancies to nanometer-sized pores.[125] These defect structures will likely modulate the local EDL via two possible effects: 1) the lower coordination number of carbon atoms resulting in more favorable adsorption of molecules/ions; and 2) the confinement effects in the case of nanopores. Similar to step edge systems, an integrated 3D-AFM and MD method would enable the determination of the accurate 3D EDL structure at these heterogeneous sites.

Another type of commonly occurring heterogeneous site at carbon electrodes is heteroatoms (oxygen, nitrogen, fluorine, etc.), which are either adventitious or intentionally introduced to tailor the surface chemistry and electronic structure of the electrodes and improve the efficiency of supercapacitors, electrocatalysis, and batteries.[126–131] Various processes have been used to incorporate the dopants into the host matrix, such as solution processing, plasma treatment, and gas annealing. In any of these processes, a large variety of defect structures are generated, including various vacancies and dopant configurations (e.g. substitutional and adsorbed). Therefore, a thorough mechanistic understanding of their interfaces with liquid electrolytes will require the determination of atomic-scale heterogeneous structures. Such investigations may also begin with plasma treatments to generate dopants in HOPG electrodes, including $O_2$, $N_2$, and $CF_4$ plasma to induce oxygen, nitrogen, and fluorine doping, respectively. Considering the high complexity of these electrodes, ex situ characterization will be necessary to first determine the overall presence and distribution of dopants. Methods may include Raman spectroscopy, X-ray photoelectron spectroscopy (XPS), scanning transmission electron microscopy (STEM) and electron energy loss spectroscopy (EELS). After the dopant structure is determined, an integrated 3D-AFM and MD investigation on a series of different dopant sites, followed by statistically significant correlation analysis of the ex situ and in situ results would enable a comprehensive determination of the EDL structure at each heterogeneous site.

**5.3 Effect of solvation structure on the activity of single-atom catalysts.** The capability to determine the exact 3D atomic-scale structure of electrode-electrolyte interfaces would have far-reaching impacts beyond energy storage. One example where atomistic solvation structure plays a critical role is electrocatalysis. According to the well-established Sabatier principle, the electrocatalytic activity depends crucially on the binding energy of the reactant molecules and intermediate species at the active sites of the catalyst.[132,133] Since the solvent molecules surrounding the active sites can strongly interact with the reactant/intermediate species (through van der Waals interaction, hydrogen bonding, etc.), they can have significant impacts on the binding configuration and binding energy.[24,25,134–139] However, to date solvation energy has mostly been calculated using continuum dielectric models, which do not contain the atomic-scale interaction effects. As a result, so far the atomistic simulations are still not able to accurately capture the kinetics of many multi-step reactions such as $CO_2$ reduction and oxygen reduction. The lack of a precise atomistic solvation model is a key bottleneck preventing the predictive design of catalyst materials and electrolyte chemistry.

One of the promising next-generation catalyst system is single atom catalysts (SACs) that consist of individual metal atoms (Fe, Co, Mg, etc.) incorporated into heteroatom-doped carbon electrodes (with N, S, or O as the dopants). These catalysts have shown high activity in various electrocatalytic reactions, and in many cases the performance is close to that of Pt-group metals while the cost is much lower.[140–142] However, their structure is highly complicated, where the active sites can have many possible coordination configurations with variations in both the metal-heteroatom and heteroatom-carbon bond arrangements. As a result, full atomistic understanding of the catalytic mechanisms has not been possible to date, due to a lack of synergistic integration of experimental characterization and theoretical calculations. The integrated 3D-AFM and MD approach will be ideal for determining the local solvation structure of the active sites, and enable further DFT calculations of the reaction pathways and kinetics.

## 6. Outlook

The future of renewable energy and sustainability hinges on our capability to design systems for the optimal interconversion between chemical and electrical energy. These energy conversion processes are inevitably dynamic, requiring the synergistic participation and oftentimes continuous evolution of electrodes, electrolytes, and their interfaces. Therefore, scientific understanding of the materials and chemical structure during the operando processes will offer the crucial guiding principles for the device and systems design and optimization. However, existing efforts on in situ characterization and computation have been heavily focused on the structure of solid electrodes and their very surface, while the "liquid side of things" have oftentimes been conveniently treated as dielectric continuum for simplicity. By demonstrating that the liquid electrolytes are structured within at least 1–2 nm from the solid surface, and that such structures are partially mobile, highly heterogeneous and different from the bulk liquid, we hope to raise the awareness that continuum dielectric approximations of interfacial liquid are way oversimplified and can lead to wrong understandings of energy conversion mechanisms. Bulk and interfacial liquid configuration is oftentimes at least equally important, if not more, compared to the solid's bulk and surface structure.

In contrast to the significant developments in not only the experimental imaging, diffraction, and spectroscopy tools but also the DFT modeling to quantify solid structures in the past few decades, our capability to elucidate the liquid composition and distribution has significantly lagged behind. Fortunately, the developments of 3D-AFM and its potential integration with MD simulations offer promising pathways to advance our understanding of interfacial liquid structures. Compared to the already mature high-resolution imaging methods, such as TEM and super-resolution optical microscopy, 3D-AFM is still in the nascent stage of development. It took half a century for TEM to achieve a lattice resolution of 0.2–0.3 nm since its first invention. Yet for 3D-AFM, the first images already reveal angstrom scale liquid layers. We believe the future is bright for 3D-AFM and potentially further improved modes of scanning probe imaging to shed light on the largely hidden liquid structures.

**Associated Content**

**Notes**

The authors declare no competing financial interest.

**Biographies**

**Lalith Krishna Samanth Bonagiri** earned his bachelor's degree in Mechanical Engineering from the Indian Institute of Technology Madras in 2019. He is currently pursuing a Ph.D. in the Department of Mechanical Science and Engineering at the University of Illinois Urbana-Champaign. His research topic is on advancing atomic force microscopy (AFM) for imaging electrical double layers. Additionally, he is integrating deep learning techniques to enhance AFM imaging capabilities.

**Amir Farokh Payam** is an associate professor (senior lecturer) at Ulster university. His research focuses on theoretical and experimental development of atomic force microscopy systems, nano/micro-resonators, solid-liquid and materials characterization and MEMS/NEMS.

**Narayana Aluru** is a professor in the Department of Mechanical Engineering and a core faculty in the Oden Institute for Computational Engineering and Sciences at the University of Texas at Austin. His research focuses on the development of multiscale methods and using them to probe physics of solid-liquid interfaces, nanofluidics, and nanomaterials.

**Yingjie Zhang** is an assistant professor at the University of Illinois Urbana-Champaign. His research focuses on in situ characterization of solid-liquid interfaces, electrocatalysis, and chemical imaging of biological cells.

**Acknowledgments**

L.K.S.B. and Y.Z. acknowledge support from the National Science Foundation under Grant No. 2137147. A.F.P. acknowledges the support by the Department for Economy, Northern Ireland through US-Ireland R&D partnership grant No. USI 186. N.R.A. acknowledges support from the National Science Foundation under Grant No. 2137157. The experiments were performed in part in the Carl R. Woese Institute for Genomic Biology and in the Materials Research Laboratory at the University of Illinois.